\documentclass[aps,groupedaddress,superscriptaddress,amsmath,amssymb,showpacs,preprint]{revtex4-1}
\usepackage{graphicx} 
\usepackage{dcolumn} 
\usepackage{bm}
\usepackage{epsfig}
\usepackage{subfigure}
\usepackage{color}

\begin{document}
\title{Hydrodynamic Model for Conductivity in Graphene}

\author{M. Mendoza}\email{Correspondence and requests for materials
  should be addressed to M. Mendoza (millmen@gmail.com)} \affiliation{
  ETH Z\"urich, Computational Physics for Engineering Materials,
  Institute for Building Materials, Schafmattstrasse 6, HIF, CH-8093
  Z\"urich (Switzerland)}

\author{H. J. Herrmann}\affiliation{ ETH Z\"urich, Computational
  Physics for Engineering Materials, Institute for Building Materials,
  Schafmattstrasse 6, HIF, CH-8093 Z\"urich (Switzerland)}

\author{S. Succi} \affiliation{Istituto per le Applicazioni del
  Calcolo C.N.R., Via dei Taurini, 19 00185, Rome (Italy),\\and
  Freiburg Institute for Advanced Studies, Albertstrasse, 19, D-79104,
  Freiburg, Germany}

\maketitle

Based on the recently developed picture of an electronic ideal
relativistic fluid at the Dirac point, we present an analytical
model for the conductivity in graphene that is able to describe the
linear dependence on the carrier density and the existence of a
minimum conductivity. The model treats impurities as submerged rigid
obstacles, forming a disordered medium through which graphene
electrons flow, in close analogy with classical fluid dynamics. To
describe the minimum conductivity, we take into account the
additional carrier density induced by the impurities in the sample.
The model, which predicts the conductivity as a function of the
impurity fraction of the sample, is supported by extensive
simulations for different values of ${\cal E}$, the dimensionless
strength of the electric field, and provides excellent agreement
with experimental data.

\pagebreak

Graphene has revealed a series of amazing properties, such as
ultra-high electrical conductivity \cite{natletter, Geim1}, ultra-low
shear viscosity to entropy ratio \cite{grapPRL}, as well as
exceptional structural strength, as combined with mechanical
flexibility \cite{mechanicalgrap} and optical transparency
\cite{transpgrap}. Many of these fascinating properties are due to the
fact that, consisting of literally one single carbon monolayer,
graphene represents the first instance ever of a truly two-dimensional
material (the ``ultimate flatland'' \cite{PhysToday}). Moreover, due
to the special symmetries of the honeycomb lattice, electrons in
graphene are shown to behave like an effective Dirac fluid of {\it
  massless} chiral quasi-particles, propagating at a Fermi speed of
about $c \sim 10^6$m/s \cite{grapPRL, grapPRB}. This configures
graphene as an unique, slow-relativistic electronic fluid, where many
unexpected quantum-electrodynamic phenomena can take place
\cite{QGP}. For instance, since electrons are about $300$ times slower
than photons, their mutual interaction is proportionately enhanced,
leading to an effective fine-structure constant $\alpha_{gr} =
e^2/\hbar v_F \sim 1$.  As a result of such strong interactions, it
has been recently proposed that this peculiar 2D graphene electron gas
should be characterized by an exceptionally low viscosity/entropy
ratio (near-perfect fluid), coming close to the famous AdS-CFT lower
bound conjectured for quantum-chromodynamic fluids, such as
quark-gluon plasmas \cite{QGP}.  This spawns the exciting prospect of
observing electronic pre-turbulence in graphene samples, as first
pointed out in Ref. \cite{grapPRL} and confirmed by recent numerical
simulations \cite{turbPRL}.

Some of the electrical properties of graphene are still not fully
understood, such as the linear increase of the electrical conductivity
with the number of charge carriers, the existence of a minimum
conductivity (see Ref. \cite{revDasSarma}, and reference therein), and
even the nature of the main scattering mechanism limiting the carrier
mobility \cite{prl2010}. In fact, classical transport theories, based
on short-range scattering of electrons by impurities, predict that the
electric conductivity in graphene should be independent of the carrier
density \cite{boltzmann}. Recent works in the field \cite{transport1,
  transport2} have shown that such linear dependence might be
potentially explained by treating the impurities as screened Coulomb
scatterers. Nevertheless, some measurements of the change in the
electrical conductivity upon immersion of graphene samples in
high-$\kappa$ dielectric media differ from this conclusion
\cite{scatter1,scatter2}. Here, we construct a model for describing
the electrical conductivity in graphene by using a completely
different approach, which is based on the recently developed picture
of an electronic ideal relativistic fluid at the Dirac point. We
demonstrate that, although this model is based on a semiclassical
theory (it cannot take into account all quantum effects, e.g. Landau
quantization, quantum hall effects, and quantum interference), it
captures the main factors that contribute to such linear behavior and
the appearance of a minimum conductivity.

Since the most likely relevant limiting factor for the graphene
conductivity is still subject of controversy, e.g. it can be due to
random charged impurity centers \cite{transport1} or strong neutral
defects that induce resonant scattering \cite{scatter1, scatter2}, we
will treat the impurities as hard-spheres, hindering the electron flow
(scattering electrons), similarly to the way a disordered medium does
in the context of fluid dynamics. The choice of hard-spheres is based
on the experimental results by Monteverde et al.\cite{prl2010}, which
suggested that electrons seem to collide mostly with short range
scatterers of the size of a few carbon-carbon interatomic distances,
like voids, adatoms, etc. Since the relativistic fluid approach is
derived from the quantum Boltzmann equation (QBE) for graphene
\cite{grap2PRB}, a hydrodynamic description of the conductivity can be
expected to apply as long as the QBE collision operator takes proper
account of the Coulomb interactions between electrons. Therefore, once
Coulomb interactions are included in the viscosity of the fluid, the
conductivity (which in our case, unlike viscosity, is a property of
the {\it flow} rather than of the {\it fluid}) becomes a function of
the dissipation introduced in the system by the impurities, i.e. the
electron-impurity interaction.

Here, we treat graphene as a disordered medium and develop a model for
its conductivity, as a function of the impurity density describing the
anomalous dependence of the conductivity on the carrier density and
the minimum conductivity due to the carrier density induced by the
impurities. The results are compared with experimental data yielding
very satisfactory agreement.

\section*{Results}

\subsection*{Electronic Fluid in Graphene}\label{electrographene}

Our treatment is based on the hydrodynamic equations derived by
M\"uller et al. \cite{grapPRL, grapPRB}, based on the quantum
Boltzmann equation for electrons in graphene. This analysis delivers
the value of the transport coefficients, namely the fluid shear
viscosity, which is an input parameter in our model. The hydrodynamic
approach in graphene is valid when the inelastic-scattering rate due
to electron-electron interactions dominates. This is the case at low
doping, at high temperatures, and in moderate fields
\cite{grap3PRB}. However, to neglect electron-phonon interactions, we
will have to stay at a moderately high temperature of around $100$K
\cite{electronphonon}. In absence of magnetic fields, the
quasiparticle distribution function, $f_s$, evolves according to the
quantum Boltzmann equation,
\begin{equation}\label{qbe:eq}
  \frac{\partial f_s}{\partial t} + \vec{v}_s \cdot \nabla f_s + e
  \vec{E} \cdot \nabla_{\vec{k}} f_s = - \Omega[f_s] \quad ,
\end{equation}
where $\vec{E}$ is an external electric field, $e$ the electric charge
of the electron, $\Omega[f_s]$ a collision operator that takes into
account the electron-electron interactions, $\vec{v}_s = s\; c\;
\vec{k}/|\vec{k}|$, $\vec{k}$ the wave vector, $c$ the Fermi speed
($\sim10^6$m/s), and the sign $s$, not to be confused with the entropy
density, distinguishes between electrons ($+$) and holes ($-$)
\cite{grap3PRB, jumpjuttner}.  At equilibrium, the probability
distribution function becomes the Fermi-Dirac distribution,
\begin{equation}\label{fdd:eq}
f_s (t, \vec{x}, \vec{k}) = \frac{1}{e^{(sc|\vec{k}| - \mu)/k_B T} +1} \quad , 
\end{equation}
where $\mu$ is the chemical potential and $T$ denotes the
temperature. Thus, in the hydrodynamic limit, from Eqs. \eqref{qbe:eq}
and \eqref{fdd:eq} one can derive the equations for the Dirac electron
fluid in graphene: $\partial \rho/\partial t + \nabla \cdot \left(
  \rho \vec{u} \right)=0$, for charge conservation; $\partial
\epsilon/\partial t + \nabla \cdot \left[ (\epsilon+p) \vec{u} \right]
= 0$, for energy density conservation and
\begin{equation}\label{moment}
  \rho_r \left [\frac{\partial \vec{u}}{\partial t} + \left(\vec{u}\cdot \nabla \right)\vec{u} \right ]+\nabla p + \frac{\vec{u}}{c^2} \frac{\partial p}{\partial t} - \eta \nabla^2 \vec{u} = \rho \vec{E}\quad , 
\end{equation}
for momentum conservation. Here, $\epsilon$ is the energy density, $p$
the pressure, $\rho$ the charge density, $\vec{u}$ the drift velocity,
$\rho_r \equiv (\epsilon + p)/c^2$, and $\eta$ the shear viscosity.

For the case of undoped graphene ($\mu = 0$), the presence of charge
density is due to the thermal energy and can be described by,
\begin{equation}\label{chargeth}
  \rho = \rho_{th} = e \left( \frac{k_B T}{\hbar c} \right )^2 \quad .
\end{equation}
However, when there are impurities, they can induce electric
potentials on the graphene sample and a correction due to the chemical
potential must be considered,
\begin{equation}\label{chargeinduced}
  \rho = \rho_{th} \Phi_\rho (\mu/k_B T) \quad ,
\end{equation}
where $\Phi_\rho$ is a dimensionless increasing function defined in
Ref. \cite{grapPRB}. Note that, in our analytical model, we will use
this concept in order to introduce a minimum conductivity in the
graphene sample, where the function $\Phi_\rho$ will be modeled by a
free parameter to fit the experimental data and will take into account
not only the carriers generated by the impurities but also other kind
of phenomena that could contribute to induce carrier density.

The shear viscosity $\eta$, in Eq.~\eqref{moment}, can be calculated
using
\begin{equation}\label{viscoreal}
  \eta=C_\eta \frac{M (k_B T)^2}{4 \hbar c^2 \alpha^2} \quad ,
\end{equation}
where $C_\eta \sim O(1)$ is a numerical coefficient, $\alpha =
e^2/\varepsilon \hbar c$ is the effective fine structure constant,
$\varepsilon$ the relative dielectric constant of the substrate, and
$M$ the number of species of free massless Dirac particles
\cite{grapPRL, grapPRB}. Additionally, the entropy densities can be
calculated according to the Gibbs-Duhem relation $\epsilon + p = T s$.
These equations have been derived under the assumption $|\vec{u}| <
c$, and therefore the relativistic correction term, $\propto\partial p
/\partial t$, can be neglected, so that the classical Navier-Stokes
equations are recovered.  Note that, despite the high speed of the
electrons, $|\vec{u}| \sim 0.1 c$, the Reynolds number remains
moderate, due to nano-metric size of the samples and the high {\it
  kinematic} viscosity of the electronic fluid in graphene.

\subsection*{Kinematic Viscosity}

Based on Ref.~\cite{grapPRL}, the dynamic viscosity of graphene in a
sample of linear size $L_0$, is given by Eq.~\eqref{viscoreal}. This
equation can be written in the following form:
\begin{equation}
\label{ETA}
\eta= C_\eta \frac{M}{4 \alpha^2} \; \left (\frac{k_BT}{\hbar
    \omega_f} \right)^2 \; \frac{\hbar}{L_0^2},
\end{equation}
where we have introduced the characteristic frequency
$\omega_f=c/L_0$, and by solving the appropriate quantum Boltzmann
equation, it is concluded that $C_\eta \simeq 0.449$. Eq.~\eqref{ETA}
can also be rewritten as $\eta = C_{\eta} q_f^{-2} \hbar/L_0^2$, where
$q_f \equiv \hbar \omega_f/(k_B T)$. Note that, in order for a
classical (non quantum) picture of electron fluid to apply, the energy
of excitations must be much lower than the thermal energy, i.e.  $q_f
\ll 1$, the so-called collision-dominated regime.  Taking a typical
set of parameters (in MKS units), $c=10^6$, $L_0=10^{-6}$, $T=100$K,
and $ \eta/s \sim 0.2 \hbar/k_B$, we obtain $\eta \sim 10^{-20}$.
Since the Reynolds number is dictated by the kinematic viscosity of
the fluid, $\nu$, rather than by the dynamic one, $\eta=\rho \nu$,
with no need of involving the mass density, it is of interest to
estimate the kinematic viscosity of the electron fluid in graphene.

To this purpose, we appeal to the definition of the Reynolds number as
given in Ref.\cite{grapPRL}, namely:
\begin{equation}
  Re = \frac{s/k_B}{\eta/\hbar} \frac{k_B T}{\hbar \omega_f} \frac{u_0L_0}{\nu_0} \quad ,
\end{equation}
where $\nu_0=cL_0$. By writing $Re=u_0L_0/\nu$ and equating with the
above, we obtain
\begin{equation}
  \nu = \nu_0 \frac{\hbar \omega_f}{k_B T} \frac{s/k_B}{\eta/s}
  \quad . 
\end{equation}
Using $\eta/s = 0.2 \hbar/k_B$ \cite{grapPRL} and $q_f \simeq 0.07$,
we obtain $\nu \simeq 10^{-2}$. To be noted that, in spite of its
extremely low dynamic viscosity, the {\it kinematic} viscosity of
graphene is about four orders of magnitude larger than that of
water. These four orders of magnitude are more than compensated by the
large speed of the electrons, which is ultimately responsible for the
sizeable values of the Reynolds numbers which can be achieved in
graphene samples at micron scales. For instance, by taking $u_0 = 0.1
c \sim 10^5$ m/s, for a sample of $1$ micron in length, we obtain $Re
\sim 20$ for the global sample, and about $Re\sim 0.04$ on the scale
of the impurities.

\subsection*{Analytical Model Description}\label{analymodel}

In this work, we will treat impurities as circular rigid obstacles of
diameter $d$. This choice is not arbitrary, but it is based on the
fact that some experiments \cite{prl2010,scatter1,scatter2} suggest
the the main scattering mechanism in graphene is due to strong neutral
defects, with a range shorter than the Fermi wavelength, inducing
resonant scattering. Thus, the diameter $d$ can be interpreted as the
characteristic length for the range of the interaction
electron-impurity.

\begin{figure}
  \includegraphics[scale=0.41]{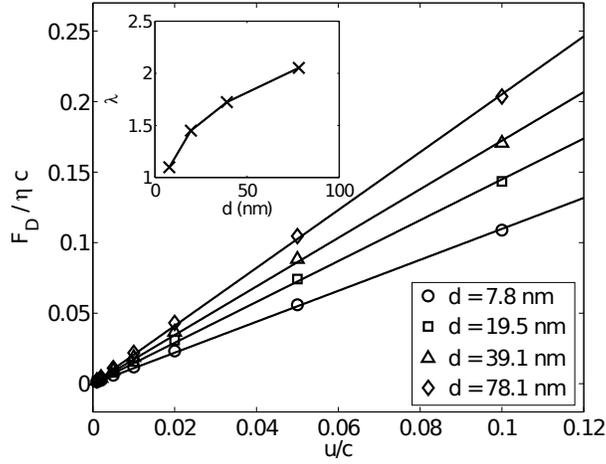}
  \caption{{\bf Drag force acting on a single impurity.} Drag force
    $F_D$ acting on a single impurity as a function of the graphene
    flow drift velocity for different impurity diameters. The solid
    lines represent the linear dependency of the drag force with the
    velocity of the fluid. In the inset, the dependence of the
    dimensionless parameter $\lambda$ on the impurity diameter is
    shown. \label{fig1}}
\end{figure}

Let us now assume that the electronic fluid moves in the $x$ direction
as a consequence of an applied electric field $E$, and $\nabla \cdot
(\rho_r \vec{u}) \simeq 0$ (incompressible limit). Therefore, we can
calculate the force $\vec{F}$ acting on a single impurity due to the
electronic flow, as $\vec{F} = \oint \tensor{\Pi} \cdot d\vec{l}$,
where $\tensor{\Pi}$ is the stress tensor defined by $\Pi_{ij} =
p\delta_{ij} + \rho_r u_i u_j + \pi_{ij}$, with the viscous tensor
$\pi_{ij} = \eta(\partial u_i/\partial x_j + \partial u_j/\partial
x_i)$, and $\vec{l}$ is a unit vector perpendicular to the impurity
circle. Here the indices $i, j = 1, 2$ denote the coordinates $x$ and
$y$.  Thus, solving the equations for the Dirac electron fluid, in the
steady state (all time derivatives are neglected), and following an
analogous procedure as in classical hydrodynamics \cite{Oseen}, we
obtain for the drag force, $F_D = F_x$,
\begin{equation}\label{stokes}
  F_D = \lambda \eta v \quad,
\end{equation}
where $\lambda$ is a dimensionless parameter that depends on the
Reynolds number. Here, $v$ is the velocity of the fluid very far from
the impurity. We first simulated single impurities with different
sizes and different fluid velocities (see Fig.~\ref{fig1}), obtaining
that a linear approximation is justified in the range of relevance to
this work. Note that, for a perfect fluid ($\nu = 0$), there is no
drag force.  However, from the point of view of the kinematic
viscosity, electrons in graphene are far from being a perfect fluid.
Therefore, we expect them to play a crucial effect on the drag force
controlling the total conductivity of the sample.
 
Let us denote by $\phi$ the impurity fraction, namely the ratio
between the area not occupied by the impurities and the total area of
the system, $\phi = 1 - N \pi d^2 / (4 A)$, with $N$ the number of
impurities in a sample of area $A$.  We can thus propose the relation
between $v$ and the total current density as $J = \phi \rho v$. Note
that $\phi \rho$ is the existent electronic charge density in the
graphene sample, since the volume fraction $1-\phi$ is occupied by the
impurities.

Let us consider a representative elementary area of the sample of
length $d_x$ in the direction of the flow and $d_y$ across it.  With
${\cal N}$ circular impurities in this area, we can write ${\cal N} =
4(1-\phi) d_x d_y/(\pi d^2)$. The total force acting on the electronic
fluid due to impurities (equal to the total force acting on the
impurities due to the fluid), is given by $F_{tot} = {\cal N}
F_D$. Here, the distance between impurities is taken sufficiently
large enough to prevent the flow close to an impurity from affecting
the flow nearby another impurity.

In order to describe correctly the physics of graphene, we need to
include in our model an extra feature. Due to the linear Dirac-Weyl
spectrum of graphene, and the non-existence of a gap between the
conduction and valence bands, the slightest amount of impurities or an
external potential will induce charge carriers in the graphene sample
\cite{transport5, Fuhrer}, see Eq.~\eqref{chargeinduced}. Thus, in our
model, the total amount of carriers induced by the impurities will be
proportional to the impurity concentration, $(1-\phi) A$, leading to
an extra carrier density in the fraction of the sample occupied by the
electronic fluid, $\phi A$.The extra carriers are then given by
$\gamma (1-\phi)/\phi$, where $\gamma$ is the proportionality constant
that characterizes the impurity-electron interaction.

Summing the forces, namely the Lorentz and drag forces, acting on the
elementary area leads to: $\phi \rho E d_x d_y + \gamma (1-\phi)E d_x
d_y/\phi - F_{tot} = 0$, and by inserting the value of $F_{tot}$, we
obtain
\begin{equation}
  F_D = \frac{\rho \pi d^2}{4} \left(\frac{\phi}{1-\phi} + \frac{\gamma}{\rho \phi} \right) E \quad .
\end{equation}
Replacing Eq.~\eqref{stokes}, taking into account that $J = \phi \rho
v$ and Ohm's law, we can identify the conductivity as:
\begin{equation}\label{conductivity}
  \sigma = \frac{\rho^2 \pi d^2}{4\eta \lambda}\left( \frac{\phi^2}{1-\phi} + \frac{\gamma}{\rho} \right)   
  = \sigma_0 \frac{\phi^2}{1-\phi} + \sigma_{min} \quad ,
\end{equation}
where we have introduced the coefficients $\sigma_0 = \rho^2 \pi
d^2/(4\eta \lambda)$, and $\sigma_{min} = \sigma_0\gamma/\rho$. This
equation represents the key result of our paper. An analogous
derivation, for fluid dynamics in disordered media, can be found in
Ref.  \cite{Rumer, Bear}. Note that $\sigma_0$ also can be written as
$\sigma_0 = n (e^2 \pi d^2 c^2/12 k_B T \nu \lambda)$, where
$n=\rho/e$ is the electronic number density and $\nu$ the kinematic
viscosity.  From this expression, we can see that the conductivity in
graphene depends linearly on the carrier density, thus implying a
constant mobility $\mu = \sigma/ne$, in agreement with experimental
observations \cite{natletter, nature2}.  In addition, our model can
also explain why the mobility remains almost constant in the range of
temperatures where $\nu \propto T^{-1}$ (see Ref.~\cite{turbPRL}), and
the presence of a minimum conductivity in graphene, second term on the
rhs of Eq.~\eqref{conductivity}, $\sigma_{min} = e\pi d^2 c^2
\gamma/12 k_B T \nu \lambda$, which is independent of the carrier and
impurity densities. Indeed, this model cannot explain, as other
theoretical models, the sublinear behavior of the conductivity for the
zero range impurity because, in that range, point defects and boundary
conditions start to be dominant. In addition, it cannot describe the
electron-phonon interaction either, since these have been excluded at
the outset.  All features above will make the object of future
extensions of this work.

% >>>>>>>>>>>>>>>>>>>>>>>>>>>>>>>>>>>>>>>>>>>>>>>>>>>>>>>>>> SS
\subsection*{Dimensionless Numbers}\label{dimeneqs}

For the numerical validation, and in order to obtain general results,
we will use dimensionless numbers. For this purpose, we can rewrite
Eq.~\eqref{moment} alternatively as
\begin{equation}\label{moment2s}
  \frac{\partial \vec{u}}{\partial t} + \left(\vec{u}\cdot \nabla \right)\vec{u} +\frac{1}{\rho_r}\nabla p + \frac{\vec{u}}{\rho_r c^2} \frac{\partial p}{\partial t} - \frac{\eta}{\rho_r} \nabla^2 \vec{u} = \frac{\rho}{\rho_r} \vec{E} , 
\end{equation}
and therefore we obtain,
\begin{equation}\label{moment3s}
  \frac{\partial \vec{u}}{\partial t} + \left(\vec{u}\cdot \nabla \right)\vec{u} +\frac{1}{\rho_r}\nabla p + \frac{\vec{u}}{\rho_r c^2} \frac{\partial p}{\partial t} - \nu \nabla^2 \vec{u} = \frac{\rho}{\rho_r} \vec{E} \quad , 
\end{equation}
where $\nu$ represents the kinematic viscosity. Let us define the
following relations: $\vec{u} = u_0 \vec{u}'$, $\vec{t} = t_0
\vec{t}'$, $(x, y, z) = L_0 (x', y', z')$, $\rho_r = \rho_{r0}
\rho'_r$, $\rho = \rho_0 \rho'$, and $\vec{E} = E_0 \vec{E}'$, where
the prime quantities are dimensionless, and $u_0$, $t_0$, $L_0$,
$\rho_{r0}$, $\rho_0$, and $E_0$ are characteristic values for the
respective physical quantities. Thus, using the state equation
$\epsilon = 2 p$, we can deduce $p = \frac{1}{3}\rho_{r0} u_0^2 p' $
and $\epsilon = \frac{2}{3} \rho_{r0} u_0^2 \epsilon'$. Replacing
these relations in Eq.~\eqref{moment3s}, multiplying by
$\frac{t_0}{u_0}$, and using the relation $u_0 = \frac{L_0}{t_0}$, we
obtain,
\begin{equation}\label{moment4s}
  \begin{aligned}
    \frac{\partial \vec{u}'}{\partial t'} &+ \left(\vec{u}'\cdot
      \nabla' \right)\vec{u}'
    +\frac{1}{3 \rho'_r}\nabla' p' \\
    &+ \frac{1}{3}\frac{u_0^2}{c^2} \vec{u}' \frac{\partial
      p'}{\partial t'} - \frac{\nu}{u_0 L_0} \nabla'^2 \vec{u}' =
    \frac{\rho_0 E_0 L_0}{\rho_{r0}
      u_0^2}\frac{\rho'}{\rho'_r}\vec{E}' .
  \end{aligned}
\end{equation}
To simplify this equation, we can equal the characteristic velocity to
the Fermi speed, $u_0 = c$. Therefore, we obtain
\begin{equation}\label{moment5s}
  \begin{aligned}
    \frac{\partial \vec{u}'}{\partial t'} &+ \left(\vec{u}'\cdot
      \nabla' \right)\vec{u}'
    +\frac{1}{3 \rho'_r}\nabla' p' \\
    &+ \frac{\vec{u}'}{3}\frac{\partial p'}{\partial t'} -
    \frac{\nu}{u_0 L_0} \nabla'^2 \vec{u}' = \frac{\rho_0 E_0
      L_0}{\rho_{r0} u_0^2}\frac{\rho'}{\rho'_r} \vec{E}' .
  \end{aligned}
\end{equation}
We can identify two characteristic dimensionless numbers. The first
one is the Reynolds number, which is, $Re = \frac{u_0 L_0}{\nu}$, and
the second one, which we call ``${\cal E}$ number'' is defined by
${\cal E} = \frac{\rho_0 E_0 L_0}{\rho_{r0} u_0^2} = \frac{\rho_0
  V_0}{\rho_{r0} u_0^2}$, where $V_0 = E_0 L_0$ is the characteristic
electrical potential. Using these definitions, we get
\begin{equation}\label{moment6s}
  \begin{aligned}
    \frac{\partial \vec{u}'}{\partial t'} &+ \left(\vec{u}'\cdot
      \nabla' \right)\vec{u}'
    +\frac{1}{3 \rho'_r}\nabla' p' \\
    &+ \frac{\vec{u}'}{3}\frac{\partial p'}{\partial t'} -
    \frac{1}{Re} \nabla'^2 \vec{u}' = {\cal E} \frac{\rho'}{\rho'_r}
    \vec{E}' .
  \end{aligned}
\end{equation}
Note that this equation is dimensionless and therefore the universal
features of the dynamics of the system are controlled only by the
numbers $Re$ and ${\cal E}$: the latter measures the strength of the
electric drive, while the former scales inversely with the dissipation
opposing this drive. For notational simplicity, we will remove primes,
leading to
\begin{equation}\label{moment6}
  \frac{\partial \vec{u}}{\partial t} + \left(\vec{u}\cdot \nabla
  \right)\vec{u} +\frac{1}{3 \rho_r}\nabla p +
  \frac{\vec{u}}{3}\frac{\partial p}{\partial t} - \frac{1}{Re}
  \nabla^2 \vec{u} = {\cal E} \frac{\rho}{\rho_r} \vec{E} ,
\end{equation}

\subsection*{Numerical Results}\label{numsim}

\begin{figure}
  \includegraphics[scale=0.41]{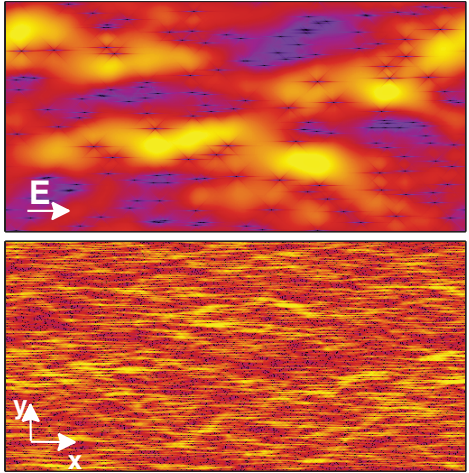}  
  \caption{{\bf Speed of the electronic flow.} Absolute value of the
    velocity in graphene with multiple impurities, for two different
    impurity fractions, $0.952$ (bottom) and $0.999$ (top). The
    electric field is applied in the $x$ direction (from left to
    right) and set up to $1.77$ V/m.\label{fig2}}
\end{figure}

Fig.~\ref{fig2} illustrates the speed of the fluid for two different
impurity densities, dark and yellow colors denoting low and high
speeds respectively.  An electric field of $1.77$ V/m was applied in
$x$ direction (from left to right). Here we can see that for high
impurity fraction (see Fig.~\ref{fig2}, top), the speed of the fluid
presents fluctuations on larger scales affecting the total
conductivity of the sample. From the calculation of the electric
current density and the electric field, we obtain the Ohm's law,
giving a slope $\sigma$.

\begin{figure}
  \includegraphics[scale=0.41]{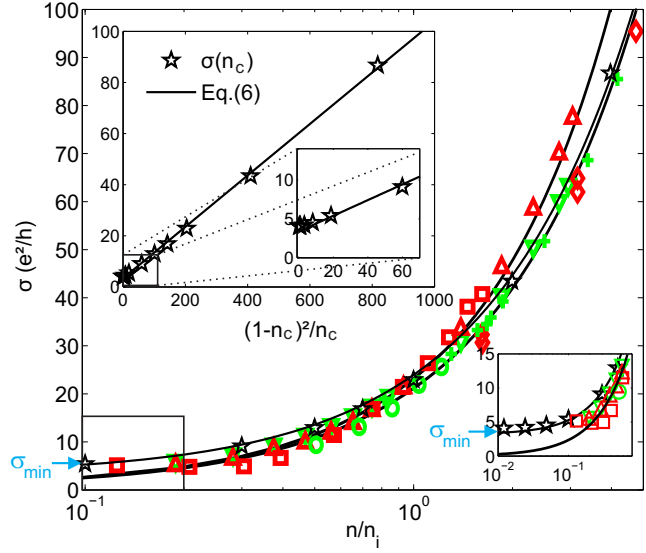}
  \caption{{\bf Comparison between our analytical model and
      experimental data.} Comparison between our results (stars) and
    experimental data for the conductivity $\sigma$, as a function of
    $n/n_i$. Data from Ref.~\cite{natletter} are represented by up and
    down triangles, from Ref.~\cite{Fuhrer} by circles and squares,
    and from Ref.~\cite{Kim} by diamonds and crosses, for electrons
    and holes respectively. Solid lines from bottom to top, theory for
    separations $\delta=0$ according to Ref.~\cite{transport1}, our
    results, and theory for $\delta=0.2 \text{nm}$ according to the
    previous reference. In the inset (top), we show the conductivity
    as a function of $\phi^2/(1-\phi)$, with an inset to observe the
    minimum conductivity. In the inset (bottom) we amplify the region
    close to the Dirac point.\label{fig3}}
\end{figure}

The conductivity $\sigma$ is calculated from the numerical slopes and
plotted as a function of the impurity fraction. The inset of
Fig.~\ref{fig3} reports the comparison between the analytical
solution, using Eqs.  \eqref{conductivity}, showing an excellent
agreement with the numerical data. For the fitting parameters, we
obtain $\sigma_0 = (9.9 \pm 0.1) \times 10^{-2} e^2/h$, and
$\sigma_{min} = 3.4 \pm 0.6$. Note that there is a difference between
our analytical model and the numerical simulations for the minimum
conductivity. This difference is due to the fact that, for high
impurity densities, the flow around one impurity starts to affect the
flow around the others, and therefore, Eq.~\eqref{conductivity} needs
some additional terms. In particular, the approximation $F_{tot} =
{\cal N} F_D$ does not hold anymore and non-linear correction terms
should be included. Thus, while the minimum conductivity given by the
analytical model is $\sim 3.4 e^2/h$, the simulations give $\sim 4
e^2/h$.  We have verified that the conductivity of graphene, as
computed in our model, does not show any appreciable dependence on the
size of the system.

In order to compare with experiments, we express the conductivity in
terms of the ratio $n/n_i$, where $n_i = ( 2.91 \times 10^{16}
\text{m}^{-2})(1-\phi)$, in our case. According to this expression and
setting $n=n_0$, we obtain that $\xi \equiv (1/\alpha) n /n_i =
(1-\phi)^{-1}$, with $\alpha = 4.85 \times 10^{-3}$.  Inserting this
result into Eq. \eqref{conductivity}, we obtain $\sigma/\sigma_0 = \xi
( 1-1/\xi)^2$.  Note that for values $\xi \gg 1$, i.e. $n/n_i \gg
\alpha = 4.85 \times 10^{-3}$, this equation tends to
\begin{equation}\label{conduc1it}
  \sigma \simeq \sigma_0 \xi + \sigma_{min} = \frac{\sigma_0}{\alpha} \frac{n}{n_i}  + \sigma_{min} \quad .
\end{equation}
This corresponds to the linear dependence obtained by different
theoretical models for graphene \cite{transport1, transport2, Fuhrer,
  transport3, transport4, transport5, transport6, transport7}.

\begin{figure}
  \includegraphics[scale=0.41]{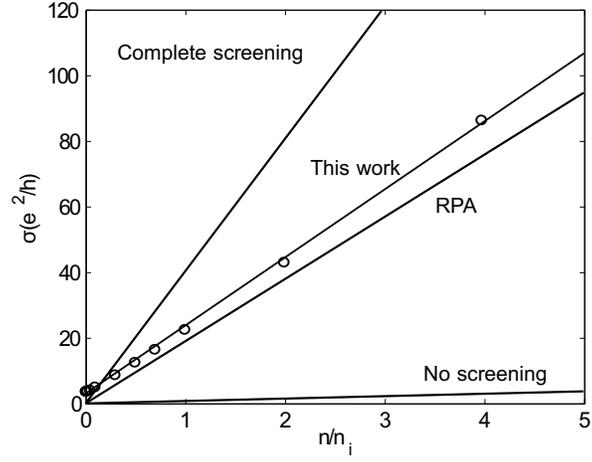}
  \caption{{\bf Comparison between analytical approaches.}
    Conductivity $\sigma$ as a function of $n/n_i$ for different types
    of scattering models \cite{transport1}. RPA is the conductivity
    calculated by using a random phase approximation with Coulomb
    scatterers. The unscreened Coulomb interaction would yield a
    conductivity smaller than the minimum value in graphene, over the
    entire range of gate voltages. \label{fig4}}
\end{figure}

In Fig.~\ref{fig3}, we see the dependence between the conductivity and
the ratio $n/n_i$, and we clearly observe the prediction for the
minimum conductivity of our model. The experimental data have been
taken from Refs.~\cite{natletter, Fuhrer, Kim}, and compared with the
results of the present work, showing good agreement.  In
Fig.~\ref{fig3}, we also compare with the model proposed by Hwang et
al.  \cite{transport1}, where the impurities are located in a plane
(substrate) parallel to the layer of graphene, with a separation
$\delta$ between the layers.  In Fig.~\ref{fig4}, we compare our
results with Coulomb impurity charges in random phase approximation
(RPA) \cite{transport1}. In the RPA model, the Boltzmann transport
equation is used with impurities that are located randomly in the
graphene sample. Our model shows good agreement in the slope with the
RPA model, however, we achieve higher values due to the shift made by
the minimum conductivity.

\section*{Discussion}

We have developed an analytical model which accounts for a linear
behavior of the conductivity with the electron density $n$, as well as
with the ratio $n/n_i$, in the limit $n/n_i \gg 4.5 \times 10^{-3}$.
In addition, it can also model the minimum conductivity in graphene as
a consequence of the carrier density induced by the presence of
impurities. Our model is based on a hydrodynamic description of
electron flow in graphene, whereby Coulomb interactions are included
through the viscosity of the electron fluid, and is valid in the
collision-dominated regime. In this model, the impurities are treated
as hard-sphere obstacles submerged on the electronic fluid, based on
the fact that some experiments \cite{prl2010, scatter1, scatter2}
suggest that strong short-range neutral scatterers are the main
scattering mechanism in graphene. Although this idea and the one about
the long-range Coulomb scatterers are still object of controversy, the
fact that the present analytical model can account for the
conductivity of graphene suggests that indeed the short-range
scattering models might be appropriate for graphene.

This work is based on the hydrodynamic description of electrons in
graphene proposed in Ref.~\cite{grapPRL, grapPRB}, which is a model
developed for undoped graphene that neglects the electron-impurity and
electron-phonon interactions. Here we have -extended- this approach by
adding the electron-impurity interactions through a macroscopic porous
media approach. Since this approach rests on basic conservation laws,
it is supposedly very robust and independent on the validity of an
underlying quantum Boltzmann equation, so long the microscopic
interactions justify the build-up of a macroscopic viscosity (no
superconductivity or other macroscopic quantum effects of that sort).
Thus, our model is able to reproduce experimental results to a
satisfactory degree of accuracy. 

For the set of parameters investigated in the present work, linear
Ohm's law appears to apply throughout. However, based on
Ref.~\cite{turbPRL}, by increasing the size of the impurities (less
screening), non-Ohmic behavior could occur, due the onset of
pre-turbulent phenomena within the graphene sample.  It would be very
interesting to verify such possibility by future experiments, as well
as the inclusion of the electron-phonon interaction to model both,
suspended and supported samples, at higher temperatures.
 
\section*{Methods}

For the simulation, we use the hydrokinetic fluid solver proposed by
Mendoza et al.  \cite{rlbPRL, rlbPRD, rlbhupp}, adapted to
two-dimensional flow in graphene \cite{turbPRL}. The simulation was
implemented on a grid of size $256 \times 512$ cells, representing a
rectangular graphene sample of size $1.5 \times 3 \mu$m. We set up
samples with a fixed number of impurities located randomly on the
grid, each impurity covering one cell size, varying $\phi$ between
$0.4$ and $0.999$. The Dirac-quasiparticle fluid in graphene has a
kinematic viscosity $\nu = 8.57\times 10^{-3} $m$^2$/s (see
Ref.~\cite{turbPRL}), and by taking the Fermi speed $u_0 = 10^6$m/s as
a characteristic speed, we obtain a Reynolds number $Re =
350$. Equating $Re$ for both systems, in physical and numerical units,
the cell size and time step are fixed to $\delta x = 5.86 \text{nm}$
and $\delta t = 5.86 \text{fs}$. For a given temperature, $T_0 = 100
\text{K}$ in our case, we can calculate the carrier density induced by
the thermal energy with Eq.~\eqref{chargeth}, $n_0 = 1.411\times
10^{14} \text{m}^{-2}$ and therefore, using the approximate relation
$\epsilon = 2 n_0 k_B T_0$ \cite{grapPRL, grapPRB}, the energy density
$\epsilon = 3.90\times 10^{-7} \text{J}/\text{m}^2$ and the density
$\rho_{r0} = 5.84 \times 10^{-19} \text{kg}/\text{m}^2$.  Using the
electron charge, we obtain the charge density, $\rho_0 = 2.26\times
10^{-5} \text{C}/\text{m}^2$.  In numerical units, these values
correspond to $n_0 = 4.845 \times 10^{-3}$, $\epsilon = 2/3$, and
$\rho_{r0} = \rho_0 = 1.0$, where the charge of the electron is $e =
2.064\times 10^2$. Using the characteristic velocity $u_0$, we can
calculate the value of the characteristic current density $J_0 =
\rho_0 u_0 = 22.6 \text{A/m}$ or $J_0 = \rho_0 u_0 = 1.0$ in physical
and numerical units, respectively. On the other hand, to obtain
realistic values of ${\cal E}$, we use a characteristic electric field
of $E_0 = 4.41 \text{V/m}$, which in numerical units corresponds to
$E_0 = 10^{-6}$. In this work, ${\cal E}$ takes values from $10^{-5}$
to $2\times 10^{-4}$. 

To model the extra carrier density induced by the impurities, as
described in Eq.~\eqref{chargeinduced}, we introduce an extra density
charge $\Delta \rho$ localized on each impurity position. Therefore,
each impurity contributes a quantity $\rho^* = \Delta \rho\; dx^2/ A$
to the total charge of the sample, such that $\rho = \rho_{th} +
\rho^* N = \rho_{th} + \Delta \rho \; n_i$, where $n_i$ denotes the
impurity density. This linear dependence between $\rho$ and $n_i$ is
in qualitative agreement with experimental data \cite{Fuhrer}. We made
several simulations for different values of $\Delta \rho$, finding
that $\Delta \rho = 60$ leads to a minimum conductivity of
$4e^2/h$. In our analytical model, this value corresponds to $\gamma =
60$, in numerical units. The simulations ran up to $5\times 10^5$ time
steps.

\bibliographystyle{naturemag}
\bibliography{report}

\section*{Author contributions}
All authors conceived and designed the research, analyzed the data,
worked out the theory, and wrote the manuscript.

\section*{Additional information}
\textbf{Competing financial interests:} The authors declare no competing
financial interests.

\section*{Figures and tables}
\pagebreak
\newpage

\end{document}